\documentclass[prd,aps,nofootinbib,onecolumn,notitlepage,longbibliography,unsortedaddress,superscriptaddress]{revtex4-1}

\usepackage[letterpaper,
  paperwidth=215.9mm,
  paperheight=279.4mm,
   top=25.4mm,
   headsep=10mm,
   bottom=25.4mm,
   footskip=18mm,
   inner=25.4mm,
   outer=25.4mm,
]{geometry}

\usepackage[T1]{fontenc}
\usepackage[utf8]{inputenc}
\usepackage{lmodern}

\usepackage{minibox}
\usepackage{boxedminipage}
\usepackage{needspace} 

\usepackage{multirow}
\usepackage{booktabs}
\usepackage{cellspace}

\usepackage{graphicx}
\usepackage{xcolor}
\definecolor{mylinkcolor}{rgb}{0.0,0.0,0.66}

\usepackage[plainpages=false,
pdftitle={},
pdfauthor={},
pdfcreator={LaTex 2e},
pdfkeywords={}
]{hyperref}
\hypersetup{colorlinks=true,linkcolor=mylinkcolor,urlcolor=mylinkcolor,citecolor=mylinkcolor,filecolor=mylinkcolor,breaklinks=true,linktocpage=true,linktoc=all}

\usepackage{bm}
\usepackage[intlimits]{amsmath}
\usepackage{amssymb}
\usepackage{upgreek}
\usepackage{slashed}
\usepackage{units}

\newcommand{\sdist}{\kern 0.20em}

\renewcommand{\eqref}[1]{Eq.\sdist(\ref{#1})}

\makeatletter
\def\normalsize{%
   \fontsize{11}{13.2}\selectfont
   \abovedisplayskip 10\p@ \@plus2\p@ \@minus5\p@
   \belowdisplayskip \abovedisplayskip
   \abovedisplayshortskip  \abovedisplayskip
   \belowdisplayshortskip \abovedisplayskip
   \let\@listi\@listI
}%
\def\small{%
  \fontsize{10}{12}\selectfont
  \abovedisplayskip 8.5\p@ \@plus3\p@ \@minus4\p@
  \belowdisplayskip \abovedisplayskip
  \abovedisplayshortskip \z@ \@plus2\p@
  \belowdisplayshortskip 4\p@ \@plus2\p@ \@minus2\p@
  \def\@listi{%
    \leftmargin\leftmargini
    \topsep 4\p@ \@plus2\p@ \@minus2\p@
    \parsep 2\p@ \@plus\p@ \@minus\p@
    \itemsep \parsep
  }%
}%
\def\footnotesize{%
  \fontsize{10}{12}\selectfont
  \abovedisplayskip 6\p@ \@plus2\p@ \@minus4\p@
  \belowdisplayskip \abovedisplayskip
  \abovedisplayshortskip \z@ \@plus\p@
  \belowdisplayshortskip 3\p@ \@plus\p@ \@minus2\p@
  \def\@listi{%
    \leftmargin\leftmargini
    \topsep 3\p@ \@plus\p@ \@minus\p@
    \parsep 2\p@ \@plus\p@ \@minus\p@
    \itemsep \parsep
  }%
}%
\makeatother

\makeatletter                                                                 
 \def\footnoterule{\kern-6\p@    %
 \hrule \@width 2in \kern 5.7\p@}  %
\makeatother

\makeatletter 
\renewcommand{\fnum@figure}{\textbf{\textsf{Fig.\sdist\thefigure}}}
\renewcommand{\fnum@table}{\textbf{\textsf{Tab.\sdist\thefigure}}}
\makeatother

\begin{document}

\title{\textsf{MP3 White Paper 2021\\[7.5pt]{}Research Opportunities Enabled by Co-locating Multi-Petawatt Lasers with Dense Ultra-Relativistic Electron Beams\vspace*{5pt}}}

	\author{Sebastian Meuren}
	\thanks{\url{smeuren@stanford.edu}}
	\affiliation{\mbox{Stanford PULSE Institute, SLAC National Accelerator Laboratory, Menlo Park, CA 94025}}%
	\author{David A.\ Reis}%
	\thanks{\url{dreis@stanford.edu}}
	\affiliation{\mbox{Stanford PULSE Institute, SLAC National Accelerator Laboratory, Menlo Park, CA 94025}}%
	\author{Roger Blandford}%
	\affiliation{\mbox{Kavli Institute for Particle Astrophysics and Cosmology, Stanford University, Stanford, CA 94309}}%
	\author{Phil H.\ Bucksbaum}%
	\affiliation{\mbox{Stanford PULSE Institute, SLAC National Accelerator Laboratory, Menlo Park, CA 94025}}%
	\author{Nathaniel J.\ Fisch}%
	\affiliation{\mbox{Department of Astrophysical Sciences, Princeton University, Princeton, NJ 08544}}
	\author{Frederico Fiuza}%
	\affiliation{\mbox{SLAC National Accelerator Laboratory, Menlo Park, CA 94025}}%
	\author{Elias Gerstmayr}%
	\affiliation{\mbox{Stanford PULSE Institute, SLAC National Accelerator Laboratory, Menlo Park, CA 94025}}%
	\author{Siegfried Glenzer}%
	\affiliation{\mbox{SLAC National Accelerator Laboratory, Menlo Park, CA 94025}}%
	\author{Mark J.\ Hogan}%
	\affiliation{\mbox{SLAC National Accelerator Laboratory, Menlo Park, CA 94025}}%
	\author{Claudio Pellegrini}%
	\affiliation{\mbox{SLAC National Accelerator Laboratory, Menlo Park, CA 94025}}%
	\author{Michael E.\ Peskin}%
	\affiliation{\mbox{SLAC National Accelerator Laboratory, Menlo Park, CA 94025}}%
	\author{Kenan Qu}%
	\affiliation{\mbox{Department of Astrophysical Sciences, Princeton University, Princeton, NJ 08544}}%
	\author{Glen White}%
	\affiliation{\mbox{SLAC National Accelerator Laboratory, Menlo Park, CA 94025}}%
	\author{Vitaly Yakimenko\vspace*{5pt}}%
	\affiliation{\mbox{SLAC National Accelerator Laboratory, Menlo Park, CA 94025}}%
\date{\today; submitted to the HFP/QED and LAP working groups}

\enlargethispage{2\baselineskip}
\begin{abstract}
\vspace*{10pt}
\noindent{}Novel emergent phenomena are expected to occur under conditions exceeding the QED critical electric field, $\unitfrac[1.3 \times 10^{16}]{V}{cm}$, where the vacuum becomes unstable to electron-positron pair production.  The required intensity to reach this regime, $\gtrsim \unitfrac[10^{29}] {W}{cm^2}$, cannot  be achieved even with the most intense lasers now being planned/constructed without a sizeable Lorentz boost provided by interactions with ultrarelativistic particles.  
Seeded laser-laser collisions may access this strong-field QED regime at laser intensities as low as  \mbox{$\sim\unitfrac[10^{24}]{W}{cm^2}$}.  Counterpropagating e-beam--laser interactions exceed the QED critical field at still lower intensities ($\sim\unitfrac[10^{20}]{W}{cm^2}$ at $\sim \unit[10]{GeV}$).  
Novel emergent phenomena are predicted to occur in the ``QED plasma regime'', where strong-field quantum and collective plasma effects play off one another. Here the electron beam density becomes a decisive factor. Thus, the challenge is not just to exceed the QED critical field, but to do so with high quality, approaching solid-density electron beams. Even though laser wakefield accelerators (LWFA) represent a very promising research field, conventional accelerators still provide orders of magnitude higher charge densities at energies $\gtrsim \unit[10]{GeV}$. Co-location of extremely dense and highly energetic electron beams with a multi-petawatt laser system would therefore enable seminal research opportunities in high-field physics and laboratory astrophysics. 
This white paper elucidates the potential scientific impact of multi-beam capabilities that combine a multi-PW optical laser, high-energy/density electron beam, and high-intensity x\ rays and outlines how to achieve such capabilities by co-locating a 3--10 PW laser with a state-of-the-art linear accelerator. 
\end{abstract}
\maketitle
\clearpage

\textbf{\textsf{Scientific goals}}
\vspace*{0.25\baselineskip}

\noindent{}The Schwinger critical electric field $E_{\text{cr}} \approx \unitfrac[1.3 \times 10^{16}]{V}{cm}$, where the quantum vacuum becomes unstable, can be reached by combining a ultra-intense optical and/or x-ray laser with a high-energy electron beam in a single facility  \cite{meuren_seminal_2020}.  This opens scientific opportunities to study fundamental questions in QED, novel plasma physics, and properties of extreme astrophysical environments. 

In this white paper, we focus on the specific opportunities made available by co-locating  a $\unit[3-10]{PW}$ laser with a $\unit[30]{GeV}$, $\sim \unit[10^{21}]{cm^{-3}}$ density electron beam. The electron beam  could be provided at SLAC, by combining the FACET accelerator test facility with the adjacent LCLS copper  LINAC using new beam compression techniques being explored at FACET-II \cite{yakimenko_facet-ii_2019}, and brought to an experimental hall where interaction with the beam of a multi-PW laser would be possible. 

Such a facility, with an intense electron beam interacting with an intense laser beam, brings two important advantages to the program of studying QED in the critical field regime.   First, such a facility would give the highest value available with current technology of the most important figure of merit for these experiments --- the value of the laser field as experienced in the frame of the high-energy electrons.   The parameters in the previous paragraph lead to field values an order of magnitude larger than the Schwinger criterion. 
Further, the density of the beam and thus the produced plasma is such that one enters the QED plasma regime, in which an electron initiates the production of a large number of electron-positron pairs that are not only produced coherently, but also display collective effects that can be rather easy to observe~\cite{qu_observing_2020}. In this regime, collective plasma and strong-field QED effects co-exist~\cite{uzdensky_plasma_2014}, and new processes appear such as coherent recollision of electron-positron pairs~\cite{meuren_high-energy_2015}. 

The large Lorentz boost of the $\unit[30]{GeV}$ electron beam significantly softens the intensity requirements to the $\unitfrac[10^{22}]{W}{cm^2}$-scale, implying that the produced pair plasma has a much lower relativistic gamma factor. This reduces the density requirements to see collective effects and therefore solves the coupled production-observation problem. As a result, this facility would make it possible not only to test the basic QED processes in the high-field region but also to study a new and unique regime of plasma physics.

This QED plasma regime has its own importance, but it also plays a role in systems of great interest in nature. One of the most enigmatic objects studied in astrophysics, the magnetar, is a neutron stars with a surface magnetic field as high as $\sim{}10^2 B_{\text{cr}}$, where $B_{\text{cr}} \approx \unit[4 \times 10^{9}]{T}$ is the Schwinger critical magnetic field \cite{kaspi_magnetars_2017}. So far, magnetars represent the only class of active galaxies  that are confirmed to be responsible for Fast Radio Bursts (FRBs) \cite{witze_astronomers_2020}.  QED plasmas are also expected to be produced in the interaction regions of high energy particle colliders \cite{yakimenko_prospect_2019,esberg_strong_2014,yokoya_beam-beam_1992}.  To understand these phenomena, we need codes that account quantitatively for the behavior of electron-positron plasmas as they pass from more familiar settings to the extreme-field regime.  This requires control of the experimental conditions, and the ability to adjust the beam parameters continuously.  This is the second advantage of co-locating a high-power laser with an electron beam from a LINAC.   The electron beam is well-defined and is characterized by well-understood diagnostics.   The transition from the low-field to the high-field regime is achieved by raising the electron beam energy systematically.

These advantages of co-location contrast with the situation for facilities with high-power optical lasers only.  It is very difficult to probe the fully nonperturbative sector of QED \cite{mironov_resummation_2020} with optical lasers, since reaching this regime requires $\sim\unit[10]{attoseconds}$ pulse durations to mitigate radiative energy losses \cite{baumann_probing_2019,blackburn_reaching_2019}. Having intense x\ rays present in such a facility could therefore be highly advantageous for studying matter in extreme electromagnetic fields due to the high frequency (penetrating power and broad bandwidth) of the radiation. 

Laser-laser collisions \cite{bell_possibility_2008} depend crucially on the seeding process \cite{tamburini_laser-pulse-shape_2017}, which is rather difficult to control. The presence of a gas or solid target at the interaction point adds significant background to the light-by-light scattering reaction.  One can create high-energy electron beams with high-power lasers using  laser wakefield acceleration~\cite{gonsalves_petawatt_2019}.  But today  conventional RF LINACs provide considerably higher particle energies, lower energy spread, and higher charge density \cite{meuren_seminal_2020,yakimenko_facet-ii_2019}, in addition to the advantages in terms of experimental control discussed above \cite{cole_experimental_2018,poder_evidence_2018}.

The multi-beam facility that we are discussing will also enable a broad physics program in topics outside strong-field QED.   These topics include advanced accelerator development, attosecond atomic and solid-state physics, and high energy density  and inertial confinement physics.   We will discuss these topics in a separate section toward the end of the white paper.

\vspace*{0.25\baselineskip}
\textbf{\textsf{Tools required}} 

\noindent{}The research opportunities presented above all make use of co-location of different combinations of three major tools: a $\unit[3-10]{PW}$ laser, a $\unit[10-30]{GeV}$ ultra-relativistic and high-density ($n \gtrsim \unit[10^{21}]{cm^{-3}}$) electron beam, and multi-millijoule multi-kilovolt x-ray free-electron lasers.   

Major pieces of the technology already exist or are planned at this moment, but co-location is a key feature that must be part of the planning, as well as focus on key critical machine parameters.

The required accelerator technology to produce dense electron beams at $\unit[\gtrsim 30]{GeV}$ is currently being developed actively at FACET-II at the $\unit[10]{GeV}$-scale \cite{yakimenko_facet-ii_2019}. The goal of 30~GeV in electron energy can be achieved at SLAC  with conventional RF acceleration if the FACET-II LINAC is connected to the current LCLS copper LINAC.   Higher electron energies can also be achieved via beam- and laser-driven plasma acceleration. 


Multiple $\unit[10]{PW}$-scale laser systems have recently been commissioned, e.g., ELI-Beamlines and ELI-NP \cite{wills_eli_2020,gales_extreme_2018,weber_p3_2017}, and more are planned. Exceeding the $\unit[10]{PW}$-scale requires coherent combination of multiple laser pulses and/or new technology, therefore this pathway is much more risky \cite{danson_petawatt_2019}.




Today, the LCLS generates x-ray pulses with a peak power of $\unit[100]{GW}$ or more and a pulse duration of about $\unit[10-20]{fs}$. The x\ rays can be focused to a spot size of $\unit[50]{nm}$ rms radius, giving at the focus an intensity larger than $\unitfrac[10^{21}]{W}{cm^2}$ and an electric field of about $\unitfrac[5\times{}10^{13}]{V}{m}$. Several recent studies have shown that it is possible to increase the peak power by two to three times using the existing system. It has also been shown that using a new superconducting undulator the power level can be increased by one order of magnitude to several TW, pushing the power density and electric field at the focus to over $\unitfrac[10^{23}]{W}{cm^2}$ and $\unitfrac[1.5\times{}10^{14}]{V}{m}$ \cite{emma_high_2016}. When backscattered on a $\unit[15]{GeV}$ electron beam the electric field $E^*$ seen by the electrons is enhanced by the beam relativistic factor, $\gamma$, to $E^*\sim\unitfrac[5\times{}10^{19}]{V}{m}$, well above the Schwinger critical electrical field, allowing the exploration of QED in a region with $E^* \gtrsim E_{\text{cr}}$ and dimensionless vector potential of order one, a regime not accessible using petawatt optical laser.

\vspace*{0.5\baselineskip}
\textbf{\textsf{Scientific impact in the area of Strong-Field QED}}
\vspace*{0.25\baselineskip}

\textit{QED plasmas:} The complex interplay between strong-field quantum and collective plasma effects in the ``QED plasma regime'' renders analytical \textit{ab initio} calculation impossible. As a result, our insights rely almost exclusively on the QED-PIC methodology \cite{gonoskov_extended_2015}, which employs many approximations. Challenging existing predictions by comparing them to experimental data will be essential for improving our understanding of the QED plasma regime. 

\textit{High-Energy Astrophysics:} Co-location of a multi-petawatt laser with a high-energy, high-density electron beam would provide a \textit{novel platform for laboratory astrophysics} that will provide important insights into the most extreme plasma conditions present in our universe \cite{uzdensky_plasma_2014}, e.g., around magnetars \cite{kaspi_magnetars_2017}, which are progenitors of fast radio bursts \cite{witze_astronomers_2020}, neutron-star merging events \cite{price_producing_2006}, and potentially also black holes \cite{castelvecchi_black_2019,blandford_electromagnetic_1977}. Relativistic pair plasmas have unique properties which differ considerably from conventional plasmas. Reaching extreme pair-plasma densities and temperatures via beam-driven QED cascades enables a novel high energy-density physics (HEDP) research frontier, that is highly relevant for the emerging field of multi-messenger astronomy.

\textit{Linear Collider Physics:} The experiments carried out at such a facility will provide important insights that could lead to a paradigm change in linear collider design: the limitations of linear lepton colliders are essentially determined by beamstrahlung \cite{esberg_strong_2014,yokoya_beam-beam_1992}. A change from long and flat to short and cylindrical bunches could drastically decrease the operational costs of the collider while ensuring scalability to the multi-10 TeV regime \cite{yakimenko_prospect_2019}. They are essential for developing the science case for a future ``nonperturbative QED collider'', which can access the fully nonperturbative regime of QED \cite{mironov_resummation_2020}, e.g., by changing the final focus design of ILC.

\vspace*{0.7\baselineskip}
\textbf{\textsf{Scientific impact in areas beyond Strong-Field QED}}
\vspace*{0.25\baselineskip}

\textit{Laser Wakefield Accelerators:} The proposed capabilities will likely have a disruptive impact on LWFA. LWFA represents a very promising technology, but does not yet deliver the combination of energy, beam density, repetition rate, and stability of the facility described here. Staging has been demonstrated with low capture efficiency \cite{steinke_multistage_2016} and seeding a LWFA stage with a high-quality, well controllable electron beam would facilitate systematic studies that could solve the challenging staging problem, essential for reaching TeV-scale energies \cite{lindstrom_staging_2021,leemans_laser-driven_2009}.

\textit{X-ray science:} 
The availability of a high-quality ultra-dense high-energy electron beam would provide a setting for an x-ray free electron laser with higher x-ray and/or pulse energies than currently operating facilities. This will open a plethora of opportunities in materials science and nuclear physics, as well as attosecond and strong-field atomic and condensed-matter physics.  In combination with the multi-PW optical laser this becomes a powerful facility for high-energy density science.

\textit{High energy density:}
This high-intensity XFEL  would enable opportunities in high energy density and inertial confinement physics not possible anywhere else.   The combination of the laser for compression of materials and the XFEL for diagnostics would be a major step forward in the study of high-pressure materials such as metallic hydrogen.  Novel physics could become accessible by focusing extremely intense hard x\ rays on a deuterium target, which is predicted to enhance the d-d fusion cross section \cite{queisser_dynamically_2019}.

\vspace*{0.5\baselineskip}
\textbf{\textsf{Broader impacts}}
\vspace*{0.25\baselineskip}

 \noindent{}Megaprojects like the $\unit[100]{PW}$ ``Station of Extreme Light'' (SEL) in Shanghai \cite{cartlidge_light_2018} are exciting the public about science and therefore play a crucial role in justifying basic research. Creating a flagship high-intensity laser facility will re-establish U.S. leadership in an area where it has fallen behind in recent years \cite{NAS_2017}. It will have a big impact on outreach programs, inspire students to pursue a scientific career and attract current and future scientific leaders. 

In the light of substantial international investment into (multi-) petawatt laser infrastructure, it is very hard to stay internationally competitive with only all-optical facilities. Therefore, it is advisable to  concentrate resources and benefit from existing infrastructure. The envisioned combination of PW lasers with  high energy,  high density electron and x-ray beams will create a unique environment that enables synergies, cross-fertilization, and novel frontiers at the intersections of these research fields. Co-location of a multi-petawatt laser system with an ultra-relativistic electron beam will provide seminal, world-wide unique research opportunities that will attract the best scientists in these research fields to the U.S. 

\vspace*{1.5\baselineskip}
\textbf{\textsf{Acknowledgments}}
\vspace*{0.25\baselineskip}

\noindent{}This work was supported by the U.S.\ Department of Energy under contract number DE-AC02-76SF00515. PHB, EG, and DAR were supported by the U.S.\ Department of Energy, Office of Science, Office of Fusion Energy Sciences under award DE-SC0020076. NJF and KQ were supported by NNSA Grant No.\ DE-NA0002948. FF was supported by the U.S.\ DOE Early Career Research Program under FWP 100331. SG was supported by U.S.\ DOE Office of Science, Fusion Energy Sciences under FWP 100182.

\clearpage
\textbf{\textsf{References}}\vspace*{-0.2\baselineskip}\\

\end{document}